\documentclass[art]{adada}  

\usepackage{graphicx}
\usepackage{color}
\usepackage{amsmath}
\usepackage{booktabs}
\usepackage{array}
\usepackage[scaled]{helvet}
\usepackage{etoolbox}

\makeatletter
\patchcmd{\@maketitle}
  {\@title}
  {{\fontfamily{phv}\selectfont\@title}}
  {}{}

\patchcmd{\@maketitle}
  {\@author}
  {{\fontfamily{phv}\selectfont\@author}}
  {}{}
\makeatother

\begin{document}

\title{Chord Colourizer: A Near Real-Time System for Visualizing Musical Key}


\artwork{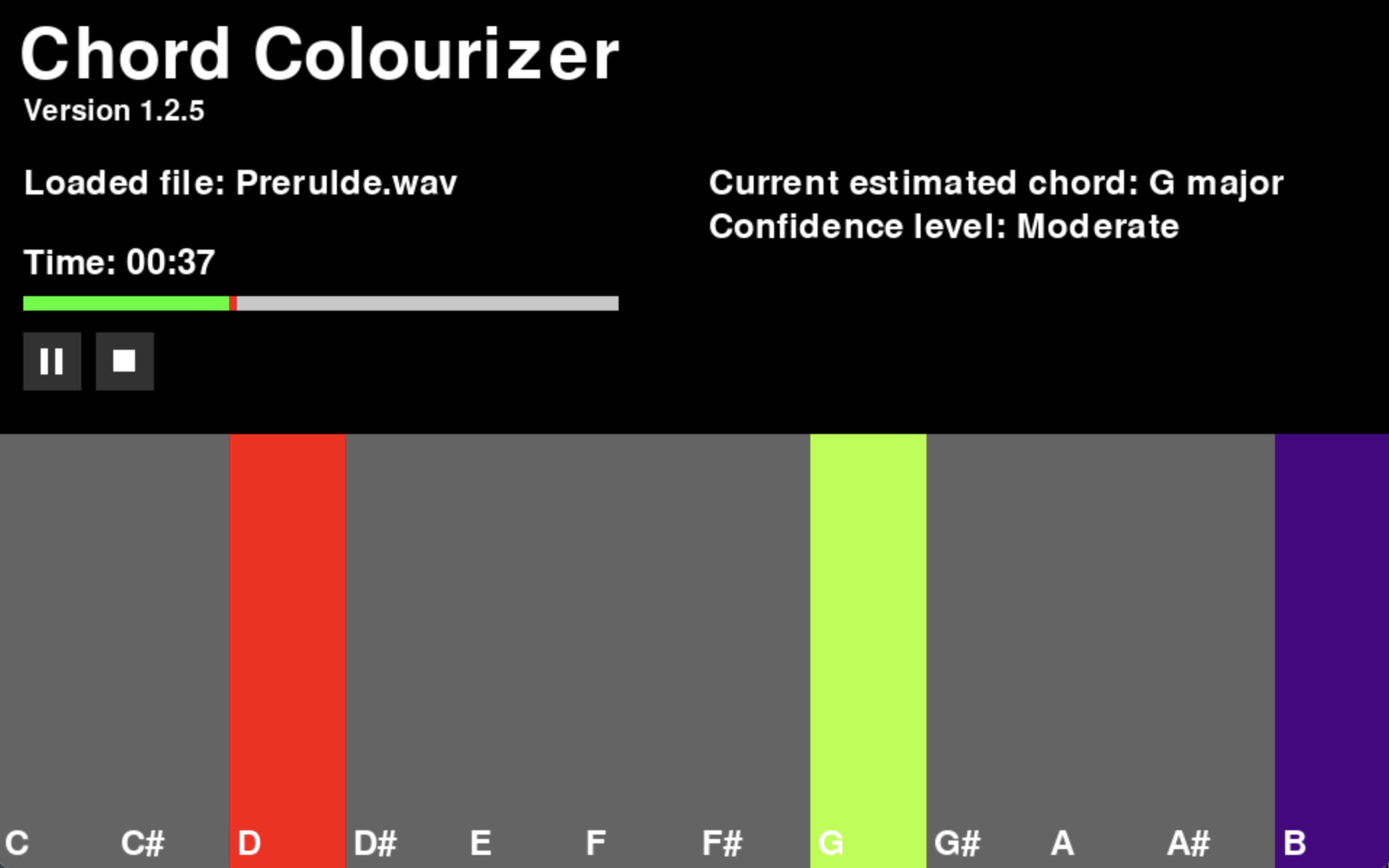}

\author{Haimes, Paul}
   {Ritsumeikan University}
   {haimes@fc.ritsumei.ac.jp}

\begin{abstract}
This paper introduces Chord Colourizer, a near real-time system that detects the musical key of an audio signal and visually represents it through a novel graphical user interface (GUI). The system assigns colours to musical notes based on Isaac Newton’s original colour wheel, preserving historical links between pitch and hue, and also integrates an Arduino-controlled LED display using 3D-printed star-shaped diffusers to offer a physical ambient media representation. The method employs Constant-Q Transform (CQT) chroma features for chord estimation and visualization, followed by threshold-based filtering and tonal enhancement to isolate the root, third, and fifth. A confidence score is computed for each detection to ensure reliability, and only chords with moderate to very strong certainty are visualized. The graphical interface dynamically updates a colour-coded keyboard layout, while the LED display provides the same colour information via spatial feedback. This multi-modal system enhances user interaction with harmonic content, offering innovative possibilities for education and artistic performance. Limitations include slight latency and the inability to detect extended chords, which future development will aim to address through refined filtering, adaptive thresholds, and support for more complex harmonies such as sevenths and augmented chords. Future work will also explore integration with alternative visualization styles, and the comparison of audio analysis libraries to improve detection speed and precision. Plans also include formal user testing to evaluate perception, usability, and cross-cultural interpretations of colour–pitch mappings.
\end{abstract}

\keywords{Music Visualization, Interactive Music Analysis, Chord Estimation}

\maketitle


\section{Introduction}
The ability to recognize and visualize musical qualities is useful for various applications, including education, performance, and analysis \cite{Muller2015}. While considerable research has been conducted on chord and key estimation \cite{Mauch2010}, the primary aim of this work is to bridge auditory and visual perception, enhancing interaction with music. This work presents an interactive, novel approach to near real-time chord representation, both on-screen and through a physical LED-based system, and is part of a larger project that also includes colour-based music generation \cite{me}. \\Inspired by Isaac Newton's original colour wheel \cite{Newton1704}, in which the spectrum of visible light was linked to musical notes by analogy. He instructed readers to ``distinguish [the colour wheel’s] circumference into seven parts... proportional to the seven musical tones"  \cite{Newton1704}. This system therefore aims to enhance the interaction between auditory and visual perception through the use of ambient media \cite{haimes,haimes2}, thus expanding on recent HCI research on colour-note mappings by providing a multi-modal interactive system \cite{Lin2024}. \\While this project builds on previous work done on cross-modal visual mappings of musical qualities such as pitch and tempo \cite{Kussner2014}, more recent research indicates that such mappings often rely on learned or metaphorical analogies rather than universal sensory correspondences, and can be shaped by modality-specific processing constraints \cite{spence2023}. Although Newton’s analogy provides a historical precedent for the sensory-artistic approach described below, this work also aims to explicitly incorporate perceptual considerations, enabling more meaningful audio-visual correspondences, and offering a framework for further exploration of alternative mappings.

\begin{figure}[thb]
\centering
\includegraphics[width=0.85\linewidth]{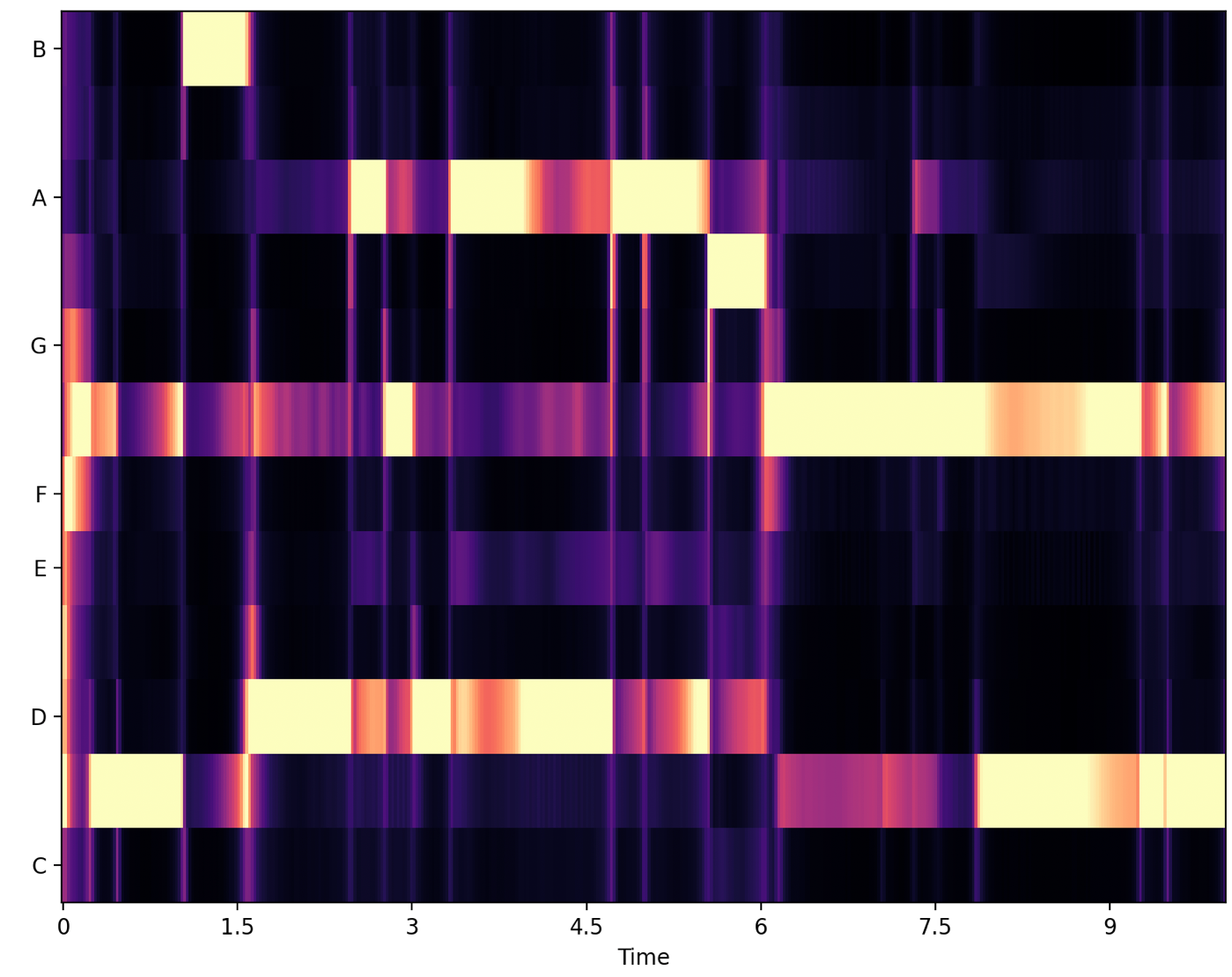}
\caption{An example chromagram time–frequency representation showing pitch classes (C to B, regardless of octave) across 10 seconds of audio}
\label{fig:chroma}
\end{figure}


\section{System design}
\subsection{Audio processing and chord detection}
The system employs the \textit{Librosa} library in Python to analyze incoming audio \cite{McFee2015}, utilizing a CQT-based chromagram to extract the most prominent musical notes (Example shown in Figure \ref{fig:chroma}). Based on this analysis, the system identifies the root note and determines the third to classify the chord, assigning a confidence score to each detection. The chord detection algorithm proceeds through several stages as described below.


\subsubsection{Chromagram averaging}

Chroma values are first calculated for audio chunks (set to 4 seconds in the current prototype). The average intensity for each pitch class is computed across time frames:

\[
\text{avgC}_p = \frac{1}{N} \sum_{f=1}^{N} C_{p,f}
\]

where:\\
- \( p \in [0,11] \): pitch class (C to B)  \\
- \( f \): frame index  \\
- \( N \): number of time frames  \\
- \( C_{p,f} \): chroma intensity of pitch \( p \) at frame \( f \)

\subsubsection{Enhancement of stronger tones}

To emphasize stronger harmonic content and reduce noise, the average chroma values are squared, boosting contrast between dominant and weak tones:
\[
\text{emphC}_p = (\text{avgC}_p)^2
\]

\subsubsection{Significance filtering}

\textbf{(1) Tone detection.} A tone is significant if:
\[
\text{emphC}_p \geq \tau_{\text{tone}} \quad (\tau_{\text{tone}} = 0.15)
\]
This provides an overview of the available tonal information.\\ \newline
\textbf{(2) Chord tone detection.} For chord analysis, a stricter threshold is applied since it's necessary to determine whether the pitch is structurally important enough to help define a chord:
\[
\text{emphC}_p \geq \tau_{\text{chord}} \quad (\tau_{\text{chord}} = 0.2)
\]

\textbf{(3) Minimum tone count.} Chord analysis proceeds only if there are at least two distinct pitch classes with strong harmonic presence:
\[
\sum_{p=0}^{11} \mathbf{1}_{[\text{emphC}_p \geq \tau_{\text{chord}}]} \geq n_{\min}
\quad (n_{\min} = 2)
\]
Here, 1 denotes the indicator function.

\subsubsection{Root note identification}

The root is the pitch class with the highest emphasized chroma value:
\[
\text{root} = \arg\max_p (\text{emphC}_p)
\]

\subsubsection{Third interval analysis}

To determine chord quality (major/minor), the minor and major thirds above the root are compared:
\[
\begin{aligned}
t_{\text{min}} &= (\text{root} + 3) \bmod 12 \\
t_{\text{maj}} &= (\text{root} + 4) \bmod 12
\end{aligned}
\]

\[
\text{qual} =
\begin{cases}
\text{Maj}, & \text{if } \text{emphC}_{t_{\text{maj}}} > \text{emphC}_{t_{\text{min}}} \\
\text{Min}, & \text{otherwise}
\end{cases}
\]
where mod 12 ensures that pitch indices map to valid notes.
\subsubsection{Fifth note computation}

The perfect fifth is automatically added to complete the triad:
\[
f_{\text{fifth}} = (\text{root} + 7) \bmod 12
\]

Finally, the detected chord is then represented as:
\[
\text{Chord} = (i_r, \text{Quality}, i_5)
\]
Note that the method described in this section assumes root-position triads only.
\subsubsection{Confidence assessment}

A confidence score is calculated from the relative energy of the thirds:
\[
c_{\text{raw}} = 
\frac{
\text{emphC}_{t_{\text{maj}}}
}{
\text{emphC}_{t_{\text{maj}}} + \text{emphC}_{t_{\text{min}}} + \varepsilon
}
\quad (\varepsilon = 10^{-6})
\]
where \( \varepsilon \) is added to avoid division by zero in cases where both the major and minor thirds are weak. \\  \newline
The raw score is scaled to a percentage:
\[
c_{\text{pct}} = \left|2 \cdot c_{\text{raw}} - 1\right| \times 100
\]


The confidence percentage is mapped to user-friendly strength ratings, as shown in Table \ref{tab:chord_strength}.

\begin{table}[h]
    \centering
    \renewcommand{\arraystretch}{1.2}
    \caption{Chord-strength classification based on confidence percentage. In the current prototype, only chords rated as ``Moderate'' and above are displayed.}
    \label{tab:chord_strength}

    \begin{tabular}{p{4cm} p{3cm}}
        \toprule
        \textbf{Confidence Percentage} & \textbf{Strength Rating} \\
        \midrule
        $c_{\text{pct}} \geq 95$     & Very Strong \\
        $80 \leq c_{\text{pct}} < 95$ & Strong \\
        $60 \leq c_{\text{pct}} < 80$ & Moderate \\
        $40 \leq c_{\text{pct}} < 60$ & Uncertain \\
        $20 \leq c_{\text{pct}} < 40$ & Weak \\
        $c_{\text{pct}} < 20$        & Very Weak \\
        \bottomrule
    \end{tabular}
\end{table}

\begin{figure}[ht]
    \centering
    \includegraphics[width=0.89\linewidth]{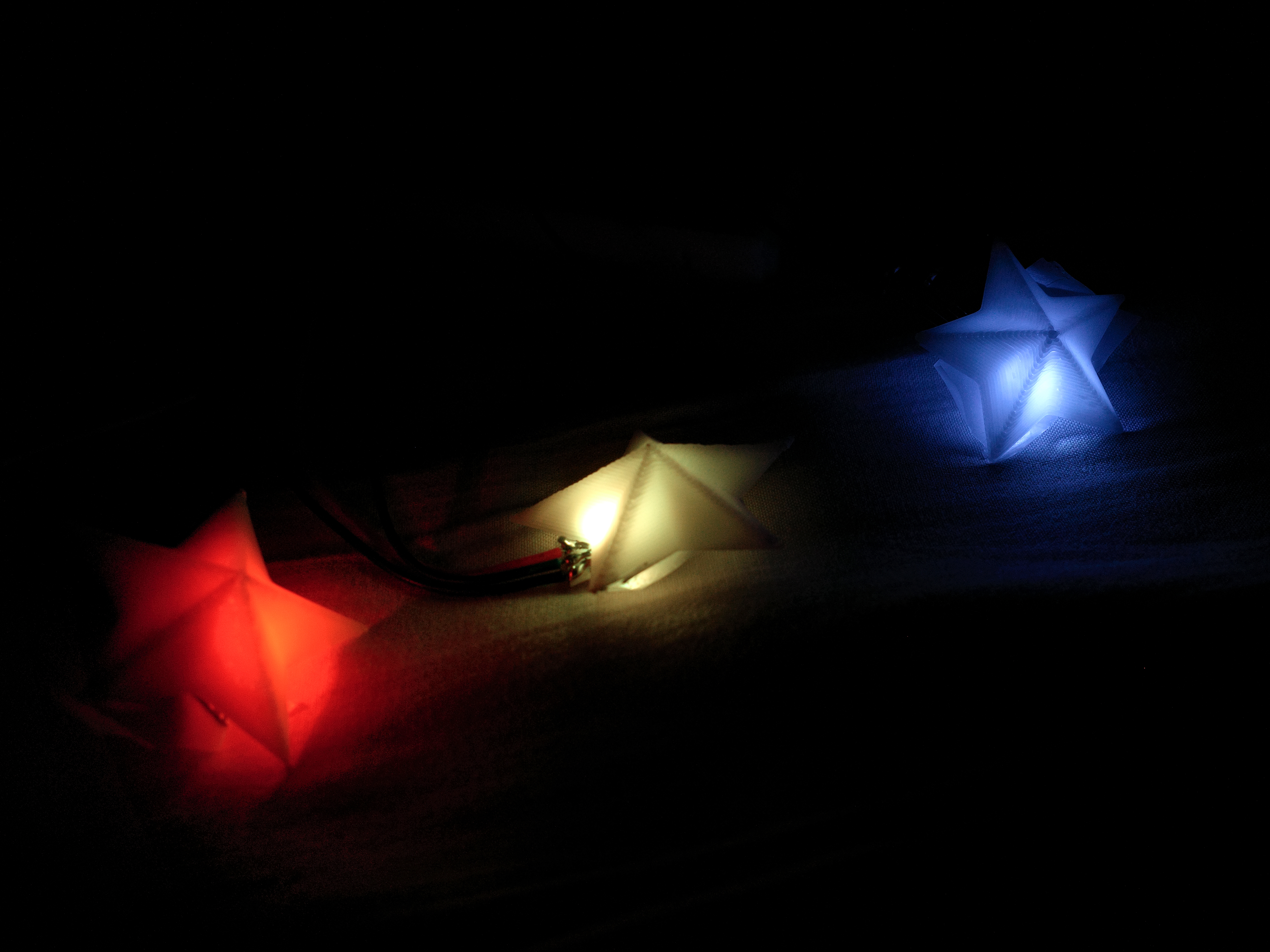}
    \caption{Prototype of the simple hardware interface,  displaying colours for the three notes of G major, which are also shown in the figure on Page 1}
    \label{fig:physical}
\end{figure}

\subsection{Visualization system}
A graphical interface, developed using \textit{Pygame}\footnote{https://www.pygame.org/docs/}, displays the detected chords in near real time. Each identified note is visually represented as a coloured rectangle on a simple keyboard layout, updating dynamically whenever a new chord is recognized. Note colours are mapped based on Newton's colour wheel, preserving the historical association between pitch and hue (Table \ref{tab:colours_notes}). For accidentals (sharps and flats), intermediary colours between Newton's primary selections are used.

The system currently includes a 4-second initialization delay to allow for stable audio analysis before beginning chord detection. During this period, the interface displays ``Initializing chord analysis..." to inform users of the system state.

\begin{figure}[thb]
\centering
\includegraphics[width=0.75\linewidth]{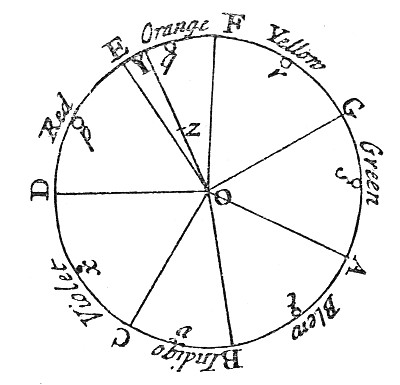}
\caption{Newton's original colour wheel c. 1704, with notes assigned to colours}
\label{fig:newt}
\end{figure}

\begin{table}[h]
    \centering
    \renewcommand{\arraystretch}{1.2}
    \caption{Notes and their corresponding colours (with RGB values), based on Newton's colour wheel (Fig.~\ref{fig:newt}).}
    \label{tab:colours_notes}

    \begin{tabular}{p{2cm} p{6cm}}
        \toprule
        \textbf{Music note} & \textbf{Colour (RGB)} \\
        \midrule
        C   & Red-violet (199, 21, 133) \\
        C\# & Violet (195, 118, 225) \\
        D   & Red (255, 0, 0) \\
        D\# & Red-orange (255, 69, 0) \\
        E   & Orange (255, 140, 0) \\
        F   & Yellow-orange (255, 165, 0) \\
        F\# & Yellow (255, 255, 0) \\
        G   & Yellow-green (173, 255, 47) \\
        G\# & Green (0, 255, 0) \\
        A   & Blue-green (83, 183, 183) \\
        A\# & Blue (0, 0, 255) \\
        B   & Indigo (75, 0, 130) \\
        \bottomrule
    \end{tabular}
\end{table}

Simultaneously, colour information is sent via serial communication to an \textit{Arduino} microcontroller\footnote{http://arduino.cc}, which operates a series of multicoloured LEDs encased in 3D-printed star-shaped diffusers (Figure \ref{fig:physical}).

\section{Results and limitations} 
Initial informal user feedback (n=5) suggests that the system effectively detects and displays chords, though a minor processing delay between audio input and visual feedback is a current challenge. So far, I have only tested the prototype with music pieces of a slower tempo (approximately 80-90 beats per minute). Future optimization efforts will focus on improving synchronization and reducing latency.

Another challenge in chord detection is the misclassification caused by transient frequencies, which can lead to incorrect chord/key recognition. At present, the system identifies the root note and major/minor third based on the strongest pitch of root-position triads only, but lacks support for extended harmonies such as major sevenths or augmented chords. The current implementation also requires a minimum of two strong tones for chord analysis, which may miss single-note passages or very sparse harmonic content.
\\
Addressing these challenges through improved filtering algorithms and support for extended chord types is a key aspect of ongoing development.

\section{Conclusion and future work}
This research presents an effective method for near real-time chord visualization, featuring both a screen-based and physical interface. The multi-stage algorithmic approach with threshold-based filtering achieves reliable chord detection while still providing just enough computational efficiency for real-time application. The system could be effectively utilized in live music settings. 

Planned enhancements include refining audio input handling, increasing chord detection precision, and experimenting with more sophisticated visualization capabilities. Future work will also explore adaptive threshold adjustment based on audio characteristics and aim to extend support for complex chord types, including sevenths and augmented chords.

Future research, building on the work of Spence and Di Stefano, may examine alternative methods of associating musical attributes with visual elements beyond Newton’s original analogy, alongside more formal user testing, which may consider cross-cultural differences in colour–note perception, among other factors.

\end{document}